\documentstyle[sprocl,epsf]{article}
\epsfverbosetrue
\def\PR #1 #2 #3 {Phys.~Rev.~{\bf #1}, #2 (#3)}
\def\PRL #1 #2 #3 {Phys.~Rev.~Lett.~{\bf #1}, #2 (#3)}
\def\PRD #1 #2 #3 {Phys.~Rev.~D~{\bf #1}, #2 (#3)}
\def\PLB #1 #2 #3 {Phys.~Lett.~{\bf B#1}, #2 (#3)}
\def\NPB #1 #2 #3 {Nucl.~Phys.~{\bf B#1}, #2 (#3)}
\def\RMP #1 #2 #3 {Rev.~Mod.~Phys.~{\bf #1}, #2 (#3)}
\def\ZPC #1 #2 #3 {Z.~Phys.~C~{\bf #1}, #2 (#3)}

\def\be{\begin{equation}}
\def\ee{\end{equation}}
\def\bea{\begin{eqnarray}}
\def\eea{\end{eqnarray}}

\begin{document}

\rightline{hep-ph/9508212}
\medskip

\title{\bf FUTURE OF TOP-QUARK PHYSICS AT FERMILAB \footnote{Presented at the
International Symposium on Particle Theory and Phenomenology, Iowa State
University, May~22-24, 1995.}}

\author{S. WILLENBROCK}

\address {Department of Physics,University of Illinois,\\ 1110 West Green
Street \\  Urbana, IL, 61801}

\maketitle\abstracts{I discuss some of the physics of the top quark
which will be explored in the near and more-distant future at the Tevatron.}

\section{Introduction}
\indent\indent Now that the existence of the top quark is firmly
established,\cite{TOP} we should begin to explore the opportunities for
top-quark physics. This is an enormous topic, as evidenced by the two-day
workshop devoted to top-quark physics following this Symposium.  In this
talk, I restrict myself to top-quark physics at Fermilab, both in the
immediate and the more-distant future. My emphasis is on top-quark physics
in the next ten years or so, during which Fermilab will have a monopoly
on the top quark.  In a final section I speculate on the role of Fermilab
during the LHC era.

The talk is divided into several subsections:
\begin{itemize}

\item Top-quark yields

\item Mass

\item Decay

\item Production

\item Not-so-rare decays

\item Speculations

\end{itemize}
The top-quark mass from combining the measured CDF and D0 values is
$179 \pm 12$ GeV.  For definiteness, I use $m_t=175$ GeV throughout this talk.

\section{Top-quark yields}

\indent\indent The machine parameters and running schedule of the
Fermilab Tevatron are given in Table~1.  Run I is now coming to a close,
and each experiment will have accumulated an integrated luminosity in
excess of 100 $pb^{-1}$ by the end of the run.  The peak luminosity
achieved thus far is about ${\cal L}=2 \times 10^{31}/cm^2/s$, impressive for
a machine that was designed for ${\cal L}=10^{30}/cm^2/s$.

\begin{table}[ht]
\begin{center}
\caption[fake]{Schedule and machine parameters for Run I and II at Fermilab.}
\bigskip
\begin{tabular}{cc}
\underline{Run I} & \underline{Run II} \\
\\
1992-1995 & 1999- \\
$\sqrt s = 1.8$ TeV & $\sqrt s = 2$ TeV \\
${\cal L}=2 \times 10^{31}/cm^2/s$ & ${\cal L}=2 \times 10^{32}/cm^2/s$ \\
$\int {\cal L} dt >$ 100 $pb^{-1}$ & $\int {\cal L} dt >$ 1 $fb^{-1}$ \\
\end{tabular}
\end{center}
\end{table}

Run II will begin in late 1998/early 1999, with a machine energy of 2 TeV.
The increase in energy is made possible by cooling the magnets to a lower
temperature, thereby allowing a higher field strength.  This increases
the top-quark production cross section by about $35\%$.

The most important change that will occur in Run II is a ten-fold
increase in luminosity, to ${\cal L}=2 \times 10^{32}/cm^2/s$.  This
will be achieved by two additions to the existing accelerator complex:
\begin{itemize}

\item Main Injector: The original Main Ring in the Tevatron collider
tunnel is a bottleneck to higher luminosity.  It will be replaced by the
Main Injector, a 120 GeV synchrotron housed in a separate tunnel, now
under construction.  The Main Injector will enable the production of many
more antiprotons, yielding a five-fold increase in luminosity.

\item Recycler: \cite{FOSTER,J} A new development within the past year is the
addition
of another element to the Main Injector project, the Recycler ring.  It is
an 8 GeV, low-field, permanent-magnet ring which will be installed in the
Main Injector tunnel.  The primary function of the Recycler is to allow
more efficient accumulation of antiprotons.  Its secondary role, from
which it takes its name, is to allow the reuse of antiprotons left over
from the previous store.  The Recycler will yield a two-fold increase in
luminosity.

\end {itemize}

There will also be a variety of detector upgrades for Run II.  One of the
most significant is an improved silicon vertex detector (SVX),
used to detect secondary vertices from $b$ quarks. This is of obvious
importance for top physics, since the top quark decays via $t\to W b$.
Both CDF and D0 will have an SVX in Run II, and they will be longer than
the existing CDF SVX, allowing for nearly $100\%$ acceptance of $b$ quarks
from top decays.  The silicon detector will also be more sophisticated,
providing a stereo view of the events.  This will increase the SVX tagging
efficiency of fiducial $b$ jets ($p_T>20$ GeV and within the SVX) from the
present value of about $40\%$ up to nearly $60\%$.  As a result of these
improvements, the fraction of top events with at least one $b$ tag will
increase from $50\%$ to $85\%$.  The fraction with two $b$ tags will
increase dramatically, from $13\%$ to $40\%$.\cite{DAN}

Taken together, the improvements in the accelerator and the detectors
will result in a dramatic increase in the potential for top-quark physics
in Run II.  In $t\bar t$ events, the final state with the most kinematic
information is
$W+4j$, where the $W$ is detected via its leptonic decay.  These events
are fully reconstructable. To reduce backgrounds, it is best to demand at
least one $b$ tag.  The number of such events is about 500/$fb^{-1}$. The
number of events with two $b$ tags, which have very small background, is
about 250/$fb^{-1}$.  Depending on the length of Run II, the integrated
luminosity delivered to each detector will be between 1 and a few
$fb^{-1}$.  Thus there will be on the order of 1000 tagged,
fully reconstructed top-quark events in
Run II, to be compared with the approximately 20 $W+4j$ single-tagged top
events
in Run I.

\section{Mass}

\indent\indent Due in part to their SVX, CDF has the best measurement of the
top-quark mass,
$m_t = 176 \pm 8 \pm 10$ GeV.  It is anticipated that the errors will be
reduced to $\pm 6 \pm 8$ GeV at the end of Run I.  If one assumes that
the error scales like the inverse of the square root of the number of
events, the error will be reduced to $\pm 3$ GeV in Run II. It may be
optimistic to assume that all the systematic errors scale like the
statistical error, so $\Delta m_t \sim 3-5$ GeV is a more conservative
prognosis.\cite{DAN}

There will also be an improved measurement of the $W$ mass in Run II, as
well as a measurement at LEP II.  An error of $\pm 50$ MeV is anticipated
from each experiment, to be compared with the current error of $\pm 180$
MeV.\cite{WMASS}  Figure~1 shows the well-known plot of $M_W$ vs. $m_t$,
with bands of constant Higgs mass. The contours show the one- and
two-sigma fits to data from LEP and SLC.  The large cross indicates the
present direct measurement of $M_W$ from CDF and UA2, and $m_t$ from CDF
and D0.  The small cross indicates the errors expected in Run II; $\Delta
M_W = 50$ MeV, $\Delta m_t = 5$ GeV, placed arbitrarily on the plot.
Note that the length of the Run II $\Delta M_W$ and $\Delta m_t$
error bars are similar.  Since the lines of
constant $m_H$ are sloped towards the horizontal, a reduction of
the uncertainty in $M_W$ yields more sensitivity to the Higgs mass than a
reduction of the uncertainty in $m_t$, a point I will return to in the
last section.

\begin{figure}[htb]
\begin{center}
\epsfxsize= 0.8\textwidth
\leavevmode
\epsfbox{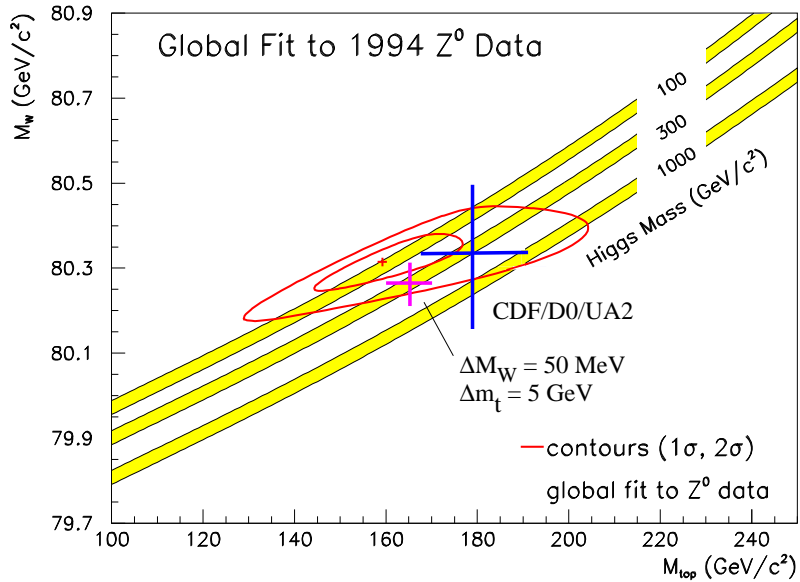}
\end{center}
\caption{$W$ mass vs. top-quark mass, with bands of constant Higgs mass.
The contours are the one- and two-sigma regions from precision LEP and SLC
data.
The large cross is the direct measurement of $M_W$ and $m_t$.  The small cross,
placed arbitrarily on the figure, is the anticipated uncertainty in $M_W$ and
$m_t$ in Run II. Adapted from Ref.~5.}
\end{figure}

There is another perspective on the top-quark mass that is interesting to
consider.  Ultimately we want to find a theory of fermion masses, and we
can ask how well we want to know the top-quark mass to help pin down
this theory.  It is reasonable to strive for a measurement of $m_t$ which
is as good (fractionally) as the best-known quark mass.  This is the $b$ mass,
which is
$\overline{m_b}(\overline{m_b}) =4.0 \pm 0.1$ GeV,\footnote{This is the running
$\overline{MS}$ mass evaluated at the quark mass.} extracted from the
Upsilon spectrum
calculated with lattice QCD.\cite{NRQCD} The top-quark mass is
already the second best-known quark mass.  Since the uncertainty in
$\overline{m_b}$ is entirely theoretical, one can anticipate that it will be
reduced
by perhaps a factor of two, corresponding to an uncertainty of $\pm 1.3\%$.
This is comparable to a 3 GeV uncertainty in the top-quark mass.
So $\Delta m_t \sim 3$ GeV is a good benchmark.

\section{Decay}

\indent\indent This section and the next are devoted to studying the
decay and production of the top quark.  To gain some perspective, I will
first ask:
\medskip

Is the Top Quark Exotic?
\medskip

\noindent There are two extreme viewpoints on this question:
\begin{itemize}

\item Yes. The top quark is much heavier than the other known fermions,
and its mass is close to the electroweak scale (e.g., the Higgs-field
vacuum-expectation value).  It seems likely that the top quark is related
to electroweak symmetry breaking.  This point of view is embodied, for
example, by top-quark-condensate models.\cite{TOPCON}

\item No. The top-quark Yukawa coupling to the Higgs field is close to
unity, a natural value.  The other known fermions have Yukawa couplings
$<<1$, which must be explained.  This point of view is embodied, for
example, by grand-unified Yukawa-matrix models.\cite{DHR}

\end{itemize}

The way to decide this issue is to study the properties of the top quark.
If the top quark is exotic, a study of its properties may reveal that fact.
One can imagine that non-standard interactions of fermions are
proportional to the fermion mass, in which case the top quark is the best
hope to discover such effects.\cite{PZ,M}

With 1000 fully-reconstructed $t\bar t$ events, the statistical accuracy
on the measurement of top-quark properties should be around $3\%$.
Including a comparable systematic error, we anticipate a knowledge of
the properties of the top quark at about the $5\%$ level at the end of
Run II.

\begin{figure}[htb]
\begin{center}
\epsfxsize= 0.25\textwidth
\leavevmode
\epsfbox{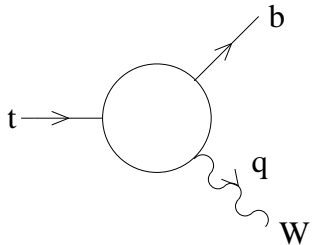}
\end{center}
\caption{The matrix element of the top-quark charged current, probed via
top-quark decay.}
\end{figure}

The weak decay of the top quark is pictured in Fig.~2.  The decay
involves the matrix element of the top-quark charged current.  Using only
Lorentz invariance, we can write down the structure of this current in
terms of four \footnote{There are two additional form factors,
\begin{displaymath}
\bar u(b)[\frac{i}{2m_t}F^3_L\sigma^{\mu\nu}P_\nu (1-\gamma_5) +
\frac{i}{2m_t}F^3_R\sigma^{\mu\nu}P_\nu (1+\gamma_5)]u(t)
\end{displaymath}
where $P=p_t-p_b$.  However, these terms do not contribute to the top-quark
decay amplitude if the $W$ boson decays to massless fermions.}
form factors \cite{KLY}
\begin{eqnarray}
\bar u(b)\Gamma^\mu u(t)& = &\frac{g}{2\sqrt 2} V_{tb}
\bar u(b)[F^1_L\gamma^\mu(1-\gamma_5) +
F^1_R\gamma^\mu(1+\gamma_5)  \nonumber \\
& & \;\;\;\;\;\;\;\;\;\; - \frac{i}{2m_t}F^2_L\sigma^{\mu\nu}q_\nu (1-\gamma_5)
-
\frac{i}{2m_t}F^2_R\sigma^{\mu\nu}q_\nu (1+\gamma_5)]u(t) \nonumber
\end{eqnarray}
where the form factors $F^{1,2}_{L,R}(q^2)$ are evaluated at $q^2=M_W^2$.
These form factors are calculable in the standard model, and are given
by \footnote{The form factors $F_1^R$ and $F_2^L$ are
zero to all orders in the standard model in the limit $m_b =0$.}
\begin{eqnarray}
F^1_L & = & 1 + {\cal O}(\alpha) + {\cal O}(\alpha_s) \nonumber \\
F^1_R & = & F^2_L\;\; =\;\; F^2_R\;\; =\;\; 0 + {\cal O}(\alpha)
+ {\cal O}(\alpha_s) \nonumber
\end{eqnarray}

An example of a measurement which provides information on the form
factors is the fraction of top decays in which the $W$ boson is
longitudinal (helicity zero) in the top-quark rest frame.\cite{KLY}
Consider the case where we set $F^1_R=F^2_L=0$, as is true in the standard
model in the limit $m_b=0$. The ratio of
the longitudinal and transverse partial widths is given by
\begin{displaymath}
\frac{\Gamma_L}{\Gamma_T}=\frac{m_t^2}{2M_W^2}
\frac{\left|1+\frac{M_W^2}{2m_t^2}\frac{F^2_R}{F^1_L}\right|^2}
{\left|1+\frac{1}{2}\frac{F^2_R}{F^1_L}\right|^2} \nonumber
\end{displaymath}
which grows quadratically with
the top-quark mass.  However, the quantity that is measured is the
branching ratio of the top quark to longitudinal $W$ bosons,
\begin{displaymath}
B(t\to W_Lb)=\Gamma_L/(\Gamma_L+\Gamma_T) \nonumber
\end{displaymath}
which is much less sensitive to the top-quark mass.  At leading order in the
standard model, $B(t\to W_Lb)= 0.70$.  A measurement of this quantity to
$5\%$ corresponds to an uncertainty in the top-quark mass of 15 GeV, much
greater than the uncertainty in a direct measurement.  Our ability to
predict this branching ratio is therefore not limited by the uncertainty
in the top-quark mass.

The form factor $F^2_R$ is non-zero in the standard model, arising dominantly
from gluon loops.\cite{L}
QCD decreases the ratio
$\Gamma_L/\Gamma_T$ by about $6\%$, which decreases the longitudinal
branching ratio by about $2\%$, to $B(t \to W_Lb) =
0.69$.\footnote{This also includes the effect of real gluon emission.\cite{L}}
A measurement of $B(t \to W_Lb)$ to $5\%$ is sensitive to a non-standard value
of $|F^2_R/F^1_L|>0.2$.

\section{Production}

\indent\indent The QCD production of top-quark pairs occurs via the
quark-antiquark annihilation and gluon-fusion processes.
The quark-antiquark annihilation process accounts for about
$80\%$ of the cross section at the Tevatron.  This process is sensitive
to the gluon coupling to top quarks,\cite{AKR,HNW} and to resonances which
might occur in this channel.\cite{HP}  The gluon-fusion process,
although suppressed, could be
greatly enhanced if there is a resonance, such as a techni-eta.\cite{AT,EL}
The measured cross section is within one sigma of the band of theoretical
predictions,\cite{LSV,BC} so there is no indication of new physics in the
production of top-quark pairs at this time.

There are two processes which produce a single top quark, rather than a
$t\bar t$ pair: the $W$-gluon-fusion process,\cite{DW,Y,EP} depicted in
Fig.~3(a),
and $q\bar q \to t\bar b$,\cite{CP,SW} shown in Fig.~3(b).  Both involve
the weak interaction, so they are suppressed relative to the QCD
production of $t\bar t$; however, this suppression is partially compensated by
the presence of only one heavy particle in the final state.  Both processes
probe the charged-current weak interaction of the top quark.
The single-top-quark production cross
sections are proportional to the square of the Cabbibo-Kobayashi-Maskawa
matrix element $V_{tb}$, which cannot be measured in top quark decays since
the top quark is so short-lived.

\begin{figure}[htb]
\begin{center}
\epsfxsize= 0.6\textwidth
\leavevmode
\epsfbox{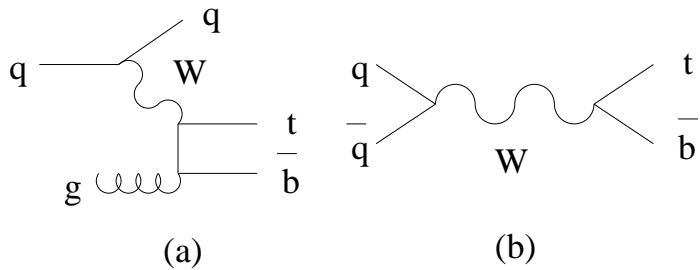}
\end{center}
\caption{Single-top-quark production at hadron colliders: (a) $W$-gluon fusion;
(b) quark-antiquark annihilation.}
\end{figure}

The single-top-quark processes lead to a final state of $Wb\bar b$ (plus
an additional jet, for $W$-gluon fusion).
The backgrounds are more serious for the single-top-quark processes than for
$t\bar t$, but they are manageable.  The dominant background is $Wb\bar
b$ from ordinary QCD/weak interactions. Fig.~4 shows a recent study of the
signal and backgrounds for $W$-gluon fusion; \cite{BH} the distribution
of the reconstructed top-quark mass is plotted for $Wjj$ events with
a single $b$ tag.\footnote{Most of the events in the signal contain one
$b$ jet plus the spectator jet from the radiation of the virtual $W$
boson.} Fig.~5 shows a similar plot for the process $q\bar q \to t\bar
b$, but with two $b$ tags.\cite{SW}  Both processes should be observed
in Run II.  The process $q\bar q \to t\bar b$ will yield a measurement of
$|V_{tb}|$ which is limited mostly by statistics; a $10\%$ measurement
may be possible in Run II.

\begin{figure}[htb]
\begin{center}
\epsfxsize= 0.5\textwidth
\leavevmode
\epsfbox{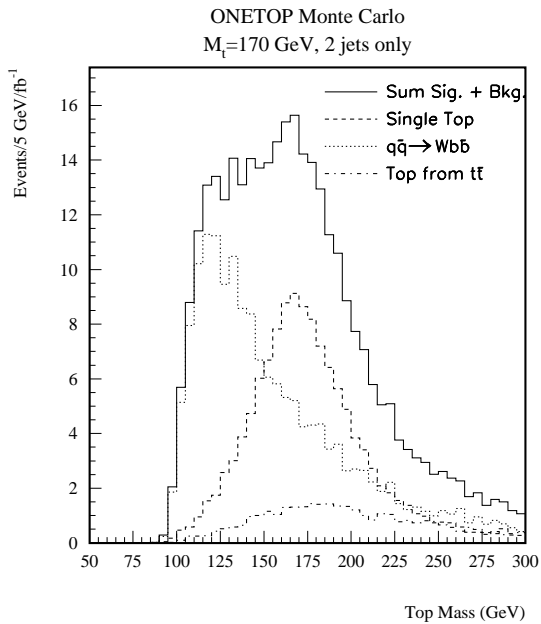}
\end{center}
\caption{Signal and backgrounds for single-top-quark production via
$W$-gluon
fusion at the Tevatron, via $Wjj$ with a single $b$ tag. From Ref.~25.}
\end{figure}

\begin{figure}[htb]
\begin{center}
\epsfxsize= 0.5\textwidth
\leavevmode
\epsfbox{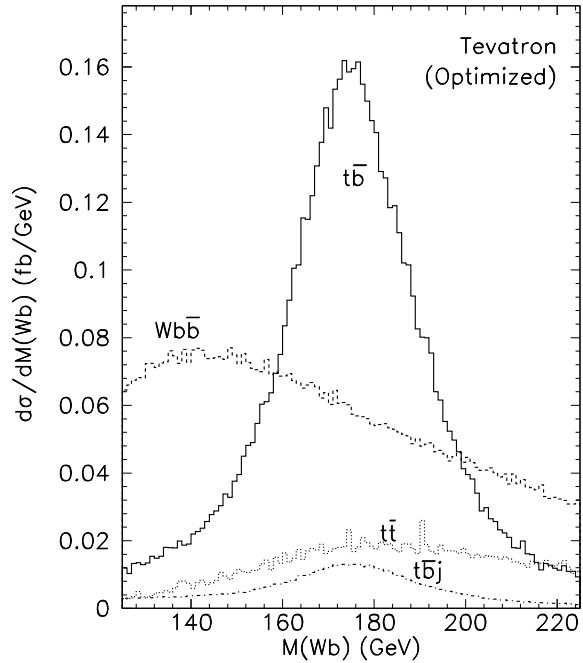}
\end{center}
\caption{Signal and backgrounds for single-top-quark production via
$q\bar q \to t\bar b$ at the Tevatron, via $Wjj$ with a double $b$ tag.
{}From Ref.~24.}
\end{figure}

\section{Not-so-rare decays}

\indent\indent The top quark can also provide a window into new physics via
its decay to new particles. In this section I discuss several decays
which could occur at the few to tens of percent level.  These decays
could be accessible even in Run I.

The decay of the top quark to a charged Higgs boson is a well-known decay
mode, and has been sought for several years.\cite{CHARGED}
The charged Higgs is sought via its decay to
$\bar\tau\nu$ or $c\bar s$.  These are the dominant decay modes in a
multi-Higgs doublet model with natural flavor conservation.\cite{GW} However,
it is conceivable that the Higgs coupling to fermions are
generation-changing, in which case the decay $H^+\to c\bar b$ could
dominate.\cite{HW} If the other top quark decays conventionally,
this would give rise to events with three $b$ quarks, which could be
distinguished by tagging all three $b$ jets.

Continuing along this line of thought, one can also imagine tree-level
flavor-changing neutral-current decays of the top quark, such as $t \to
ch^0$.\cite{HW,H}  Flavor-changing neutral currents are severly
restricted in the first two generations of fermions, but could be large
in the third generation.  The dominant decay of the neutral Higgs would
likely be $h^0 \to b\bar b$.  If the other top quark decays
conventionally, the events would again have three $b$ quarks in the final
state.

If the top squark is light, it could potentially be discovered in
top-quark decays.\cite{BDGGT,MKKW}  The decay mode
$t \to \tilde t_1 \tilde\chi^0_1$
could have a significant branching ratio ($\tilde\chi^0_1$ is the lightest
neutralino).  If it is kinematically
allowed, the top squark will decay via $\tilde t_1 \to \tilde\chi^+_1 b$;
otherwise, the loop-induced decay $\tilde t_1 \to \tilde\chi^0_1 c$ would
dominate.  The extraction of these signals from backgrounds is challenging.

\section{Speculations}

\indent\indent As discussed in the section on top-quark yields,
the Main Injector and
the Recycler will allow the Tevatron to achieve a luminosity of $2\times
10^{32}/cm^2/s$ in Run II.  One can ask if even higher luminosity can be
achieved.  The answer seems to be yes.\cite{FOSTER}  A design exists
which would achieve a luminosity of at least $10^{33}/cm^2/s$. The
main requirement to achieve this luminosity is to increase the rate of
antiproton production.  This can be attained by directing more bunches
from the Main Injector onto the antiproton-production target, a technique
called ``multibatch targeting''.\cite{FOSTER2}  The cost of this scheme is
modest in
comparison with the Main Injector, and could be in place for Run II.

A luminosity of $10^{33}/cm^2/s$ would produce about $5,000$ tagged and
fully-reconstructed top-quark pairs per year.  One can imagine pushing
the uncertainty on the top-quark mass down to $2$ GeV, and the accuracy
on the measurement of the top-quark properties down to $2-3\%$.  There
are other physics opportunities which become available as well.  The
error on the $W$ mass could potentially be pushed down to 20 MeV.  This
may be even more interesting than improving the accuracy on the top-quark
mass, as remarked in section 3.  The production of the Higgs boson in
association with a $W$ boson, followed by $H\to b\bar b$, may also become
accessible, in the mass range $m_H = 80-120$ GeV.\cite{SMW,BBD,MK}

Given the physics opportunities afforded by ${\cal L} = 10^{33}/cm^2/s$,
why don't we do it?  The stumbling block is not the accelerator, but the
detectors, which cannot operate at such a high luminosity.  Significant
detector upgrades, or perhaps a new detector, are needed to take
advantage of this luminosity. One can even imagine this occuring during the
LHC era, especially if some of the physics objectives of such a machine are
complementary to the LHC.

Can one contemplate a luminosity for a $p\bar p$ collider as high as
${\cal L} = 10^{34}/cm^2/s$, the LHC design luminosity?  There doesn't
seem to be any reason why not.  If such a luminosity can be attained, it
might remove the advantage of $pp$ colliders over that of $p\bar p$. A
$p\bar p$ collider requires only a single ring of magnets, so it can
potentially be built more economically than a $pp$ collider, which
requires either two rings, or a 2-in-1 magnet such as for the LHC.
Magnets are a significant fraction of the cost of an accelerator; they
account for roughly two thirds of the cost of the LHC, for example.
The next hadron collider after the LHC might be a return
to $p\bar p$.\cite{FOSTER}

\section*{Acknowledgements}

\indent\indent I am grateful for conversations with D.~Amidei, W.~Foster,
G.~Jackson, and T.~Liss, and for assistance from P.~Baringer, A.~Heinson,
R.~Keup, and J.~Womersley.
This work was supported in part by Department of Energy grant
DE-FG02-91ER40677.

 \clearpage

\end{document}